\documentclass[sigconf]{acmart}
\AtBeginDocument{%
  }

\usepackage{times}
\usepackage{latexsym}

\usepackage{microtype}

\usepackage{graphicx}

\usepackage{subcaption}

\usepackage{booktabs}
\usepackage{multirow}
\usepackage[linesnumbered,ruled,vlined]{algorithm2e}
\usepackage{hyperref}
\usepackage{amsmath}

\begin{document}

\title{StyleSpeech: Parameter-efficient Fine Tuning for Pre-trained Controllable Text-to-Speech}

\author{Haowei Lou}
\email{haowei.lou@unsw.edu.au}
\authornotemark[1]

\affiliation{%
  \institution{University of New South Wales}
  \city{Kensington}
  \state{NSW}
  \country{Australia}
}

\author{Helen Paik}
\affiliation{%
  \institution{University of New South Wales}
  \city{Kensington}
  \state{NSW}
  \country{Australia}}
\email{h.paik@unsw.edu.au}

\author{Wen Hu}
\affiliation{%
  \institution{University of New South Wales}
  \city{Kensington}
  \state{NSW}
  \country{Australia}}
\email{wen.hu@unsw.edu.au}

\author{Lina Yao}
\affiliation{%
  \institution{University of New South Wales}
  \city{Kensington}
  \state{NSW}
  \country{Australia}}
\email{lina.yao@unsw.edu.au}

\renewcommand{\shortauthors}{NULL et al.}

\begin{abstract}
This paper introduces StyleSpeech, a novel Text-to-Speech~(TTS) system that enhances the naturalness and accuracy of synthesized speech. Building upon existing TTS technologies, StyleSpeech incorporates a unique Style Decorator structure that enables deep learning models to simultaneously learn style and phoneme features, improving adaptability and efficiency through the principles of Lower Rank Adaptation~(LoRA). 
LoRA allows efficient adaptation of style features in pre-trained models. 
Additionally, we introduce a novel automatic evaluation metric, the LLM-Guided Mean Opinion Score (LLM-MOS), which employs large language models to offer an objective and robust protocol for automatically assessing TTS system performance.
Extensive testing on benchmark datasets shows that our approach markedly outperforms existing state-of-the-art baseline methods in producing natural, accurate, and high-quality speech.
These advancements not only pushes the boundaries of current TTS system capabilities,  but also facilitate the application of TTS system in more dynamic and specialized, such as interactive virtual assistants, adaptive audiobooks, and customized voice for gaming. Speech samples can be found in \href{https://style-speech.vercel.app/}{https://style-speech.vercel.app/}
\end{abstract}



\received{20 February 2007}
\received[revised]{12 March 2009}
\received[accepted]{5 June 2009}

\maketitle
\section{Introduction}
Text-To-Speech~(TTS) system converts written linguistic content into human-like speech, which is a crucial technology in today's digital landscape. By reducing the reliance on human speakers, TTS significantly reduces the cost of producing human speech. Make it increasingly important in applications such as smart homes~\cite{GoogleAssistant,AmazonAlexa}, robots~\cite{breazeal2001emotive}, and virtual assistant~\cite{ReadSpeaker2023,SOVA2023}. 

Research in TTS synthesis has evolved significantly over the past 30 years, transitioning from simple Hidden Markov Models~(HMM) models~\cite{masuko1996speech,yoshimura1999simultaneous,masuko1996speech} to today's sophisticated Deep Learning~(DL) approaches~\cite{wang2017tacotron,shen2018natural,ren2019fastspeech,ren2020fastspeech}. These advances strive to produce human speech with high accuracy, realism, and variability, meeting the increasing demand for high-quality speech synthesis.

Many TTS systems are purely phoneme-driving, which limits their ability to capture the natural variations found in human speech. This often results in TTS systems lacking variation and style control, making the synthesised speech less engaging and less able to adapt the nuances changes in human speech across different situations~\cite{tan2021survey, hasanabadi2023overview, seshadri2021emphasis,yao2024sr}.
\begin{figure}
    \centering
    \includegraphics[width=0.5\linewidth]{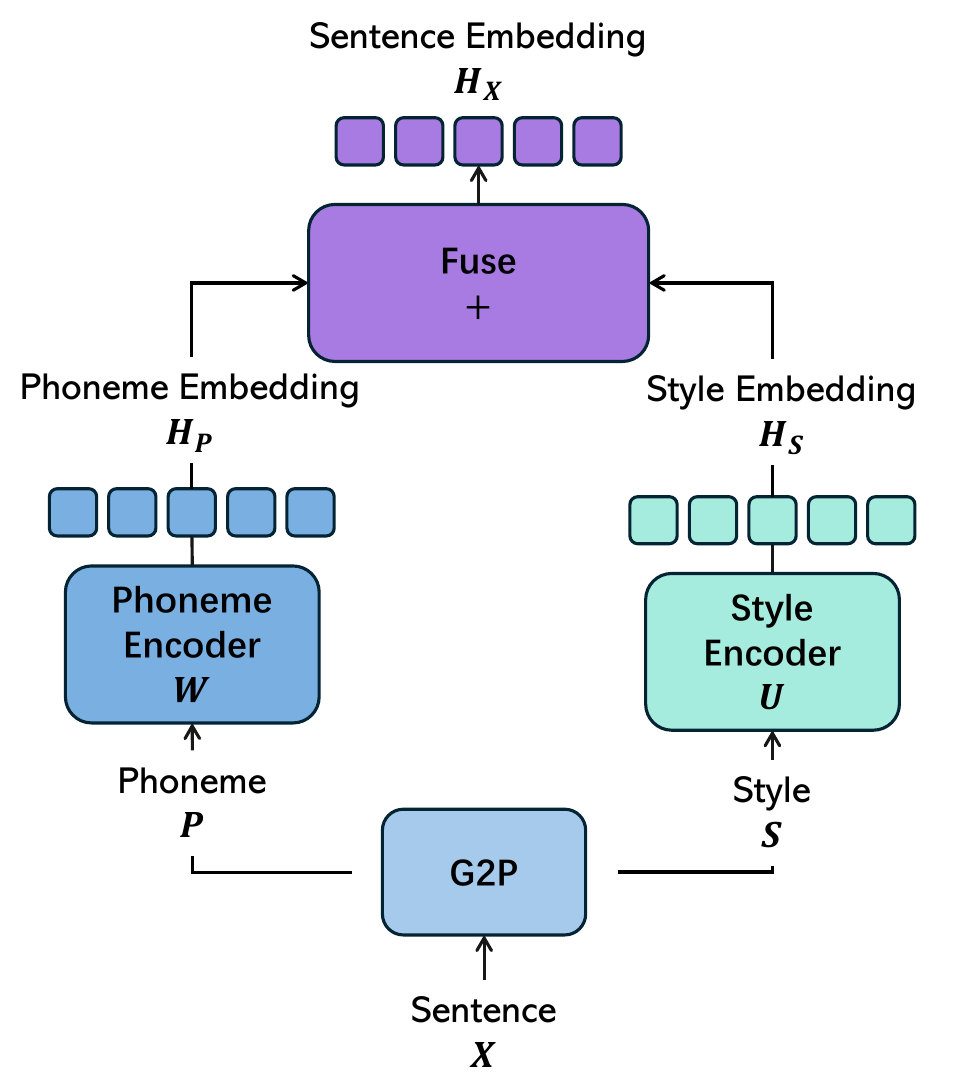}
    \caption{Style Decorator}
    \label{fig:style_decorator}
\end{figure}

Recently, FastSpeech~\cite{ren2020fastspeech} has been proposed as a state-of-the-art TTS system that integrates essential style features such as duration, pitch, and energy using feed-forward structure. 
However, this structure poses challenges, as the style encoding layer is hierarchically placed above the phoneme encoding layer within the architecture, leading the backpropagation process to prioritize updates to the style encoder’s parameters over those of the phoneme encoder.

To be more specific, phoneme embeddings are passed through the style encoding layer, where they are fused with style embeddings. The fused embeeding is treated as a unified whole during the Mel-Spectrogram decoding process. Although this method effectively incorporates style features into speech, it also leads to a significant challenge. Since the style encoding layer is positioned above the phoneme encoding layer, any updates made during back-propagation tend to prioritize the style features. This hierarchical structure causes the unique characteristics of the phoneme embeddings to blend excessively with the style features. As a result, the distinctive features of the phonemes, which are crucial for clear and accurate speech production, become diluted, leading to a loss of phonetic clarity in the synthesized speech.

Another challenge in TTS research is the absence of standardized evaluation protocols that can fairly assess the performance of a TTS system~\cite{triantafyllopoulos2023overview}. Most studies rely on the Mean Opinion Score (MOS), which depends on human listeners to evaluate the quality of synthesized speech. This method is labor-intensive and subject to the biases and variability of human perception.

To address aforementioned challenges, we introduce a new TTS framework named StyleSpeech. Building upon existing TTS systems, StyleSpeech employs a style decorator structure that adapts style features with minimal changes to existing phoneme parameters, thereby enhancing control over the synthesized speech for more accurate speech output. Additionally, we introduce the LLM-Guided Mean Opinion Score~(LLM-MOS), an innovative automatic evaluation metric that uses large language models for more objective and efficient evaluations, aiming to overcome the limitations of traditional MOS.
The main contributions of this work are:
\begin{enumerate}
    \item We propose a novel Style Decorator structure that effectively separates the training of style features from phonetic features, simplifying the process of style adaptation.
    \item  We employ the Lower Rank Adaptation~(LoRA) technique to enable efficient fine-tuning of pretrained models with minimal parameter adjustments, preserving the unique characteristics of phoneme embeddings during style adaptation.
    \item We introduce the LLM-Guided Mean Opinion Score~(LLM-MOS), a new automatic evaluation metric that leverages large language models to provide an objective and robust assessment of TTS system performance.
    \item We conducted extensive experiments on a well-known benchmark dataset and demonstrated that StyleSpeech achieves a 15\% improvement in Word Error Rate and a 12\% improvement in overall ratings compared to existing baseline models.
\end{enumerate}

\section{Related Work}
In this section, we discuss prior research that has influenced or inspired the design of the StyleSpeech framework.

Tacotron 1 and 2~\cite{shen2018natural,wang2017tacotron} are the first successful deep learning-based TTS systems that have been widely evaluated and deployed in many real-world applications. The Tacotron family of TTS systems primarily uses a sequence-to-sequence~(Seq2Seq) encoder-decoder framework to match inputs~(characters or phonemes) with the output Mel-Spectrograms with an attention module in between learns to align the input tokens with the output Mel-Spectrogram.

FastSpeech~\cite{ren2019fastspeech,ren2020fastspeech} family was subsequently proposed to address word-skipping issues in long sequence inputs and to enhance the controllability of the synthesized speech. Unlike its predecessors, FastSpeech utilises the Transformer~\cite{vaswani2017attention} structure to generate embedding sequences in parallel, which mitigates the word skipping problem and accelerates inference. FastSpeech2~\cite{ren2020fastspeech} introduces a variance adapter that offers enhanced control over style features such as duration, pitch, and energy, making the TTS output more realistic.

Lower Rank Adaptation~(LoRA) is a significant technique in Large Language Model~(LLM) research in NLP designed to fine-tune large pre-trained models efficiently by training only a small subset of parameters~\cite{hu2021lora}. This method specifically updates smaller sections of a model's parameters within crucial layers, drastically reducing computational costs. After training, these modified components are reintegrated into the original model for use during inference.
LoRA facilitates precise adaptations of models such as GPT~\cite{brown2020language} or BERT~\cite{devlin2018bert} to downstream tasks without complete retraining. The small-subset parameter ideology behind LoRA is also applicable in TTS tasks, where pre-trained TTS systems can be adapted like LLMs, treating each style adaptation as a downstream task.

AutoVC~\cite{qian2019autovc} offers a zero-shot voice style transfer technique that converts a source person's voice into a target person's style while maintaining clarity and intelligibility. It features a unique architecture where the content encoder for the source voice and the style encoder for the target voice operate in parallel. The source voice embeddings first pass through a bottleneck structure that isolates the style features from the content-related features. These isolated content features are then merged with the target style features to effectively complete the voice conversion process.

Building upon LoRA and AutoVC, we design a \textbf{Style Decorator} structure for StyleSpeech that allows deep learning models to learn style features separately and in parallel alongside phoneme features. The advantages of this structure compared with the feed-forward structure include: (1) style feature is trained as an independent module in parallel, enabling the integration of new style features without updating the entire model's parameters; (2) phonetic-related parameters are frozen during training, which preserves the uniqueness of phoneme feature; and (3) the system can integrate various types of style features as needed, enhancing the overall potential and adaptability of the TTS system. More details will be presented in Section~\ref{sec:method}.


\begin{figure*}
    \begin{subfigure}[b]{0.5\textwidth}
        \centering
        \includegraphics[width=\linewidth]{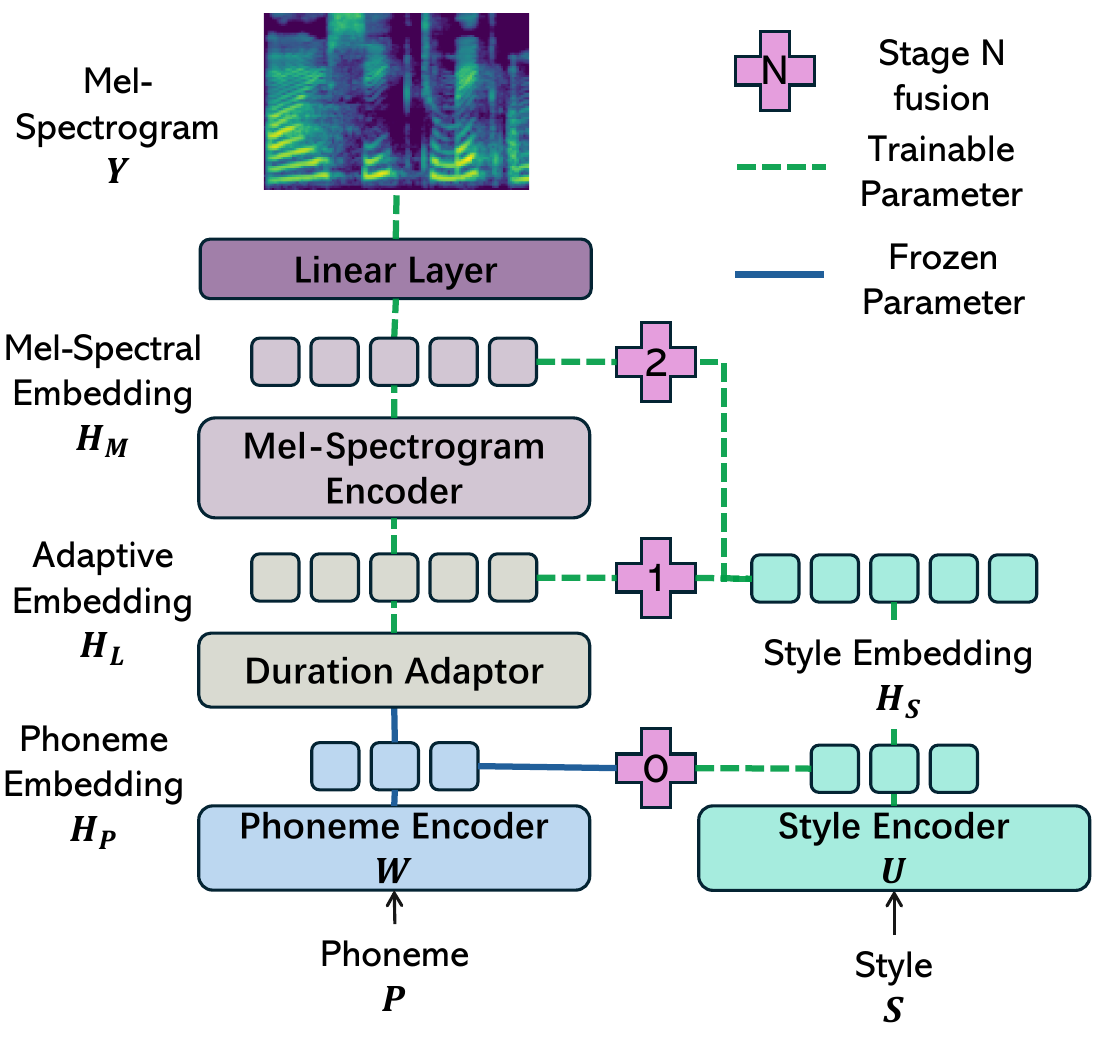}
        \caption{StyleSpeech Overview}
        \label{fig:style_speech}
    \end{subfigure}
    \begin{subfigure}[b]{0.32\textwidth}
        \begin{subfigure}{\textwidth}
            \centering
            \includegraphics[width=\linewidth]{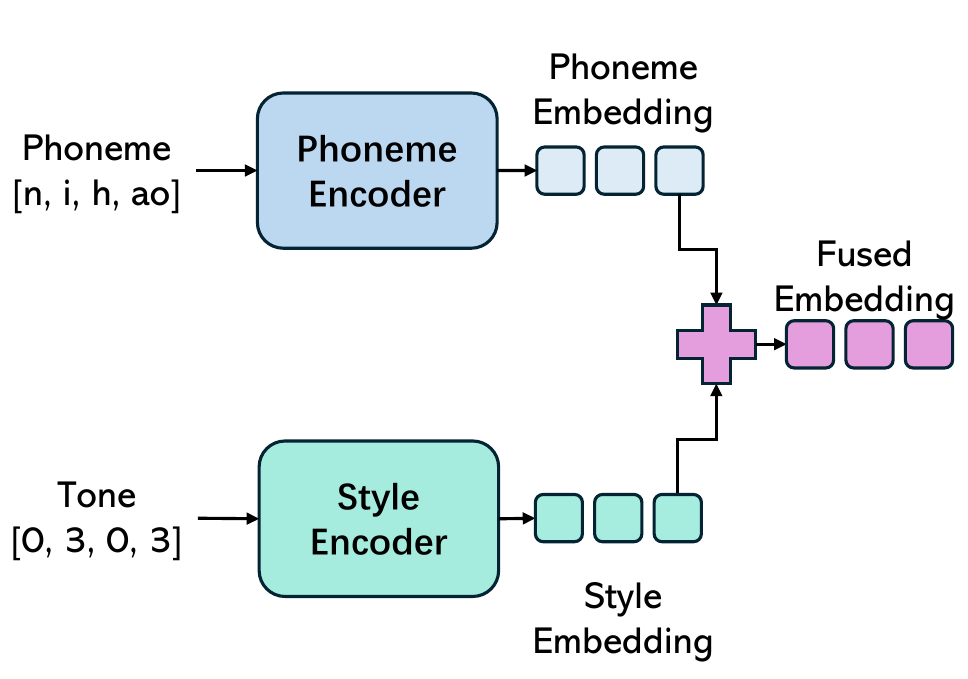}
            \caption{Fusion Process}
            \label{fig:fusion}
        \end{subfigure}      
        \begin{subfigure}{\textwidth}
            \begin{subfigure}{0.44\linewidth}
                \centering                \includegraphics[width=0.8\linewidth]{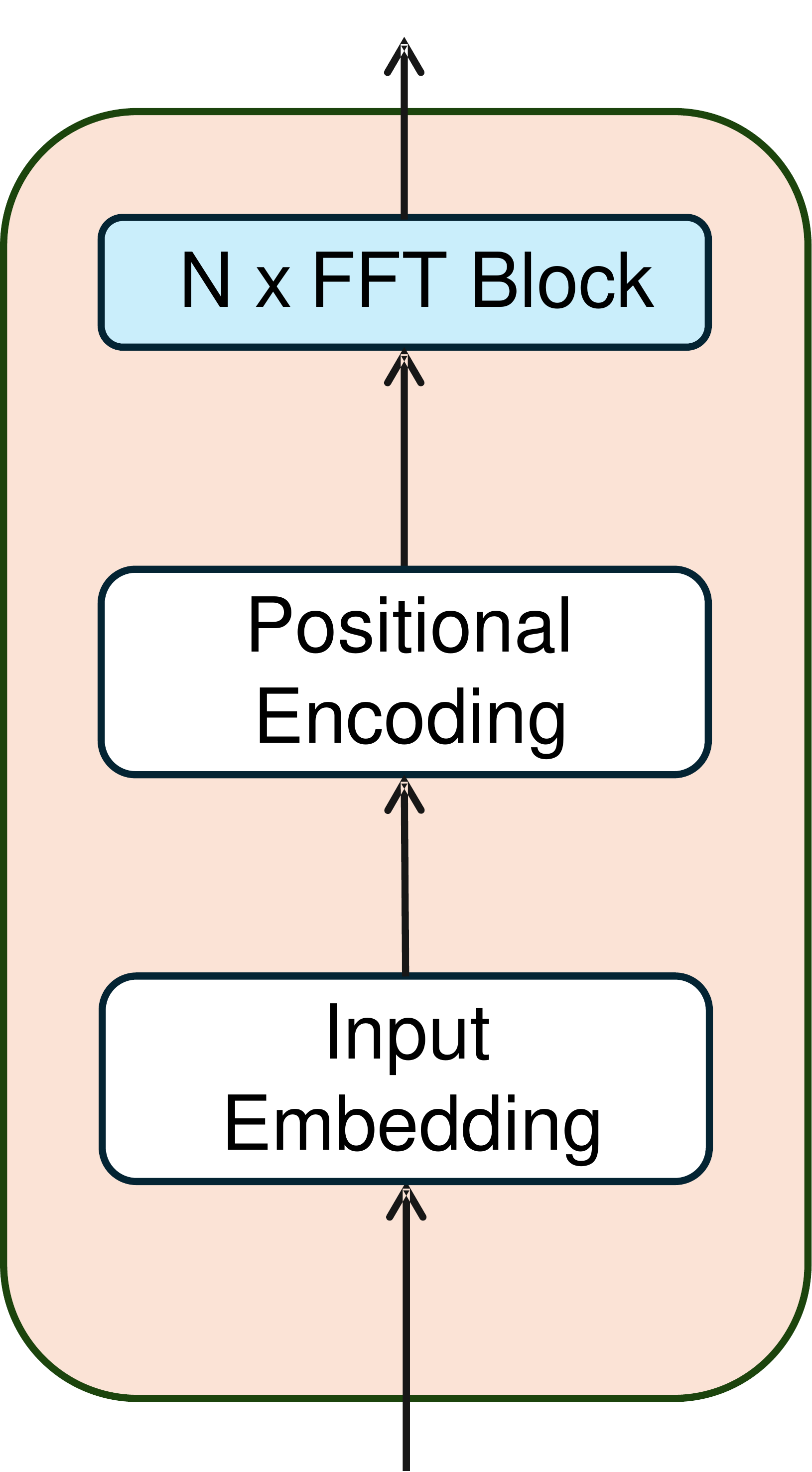}
                \caption{Acoustic Pattern Encoder}
                \label{fig:ape}
            \end{subfigure}
            \begin{subfigure}{0.44\linewidth}
                \centering
                \includegraphics[width=0.8\linewidth]{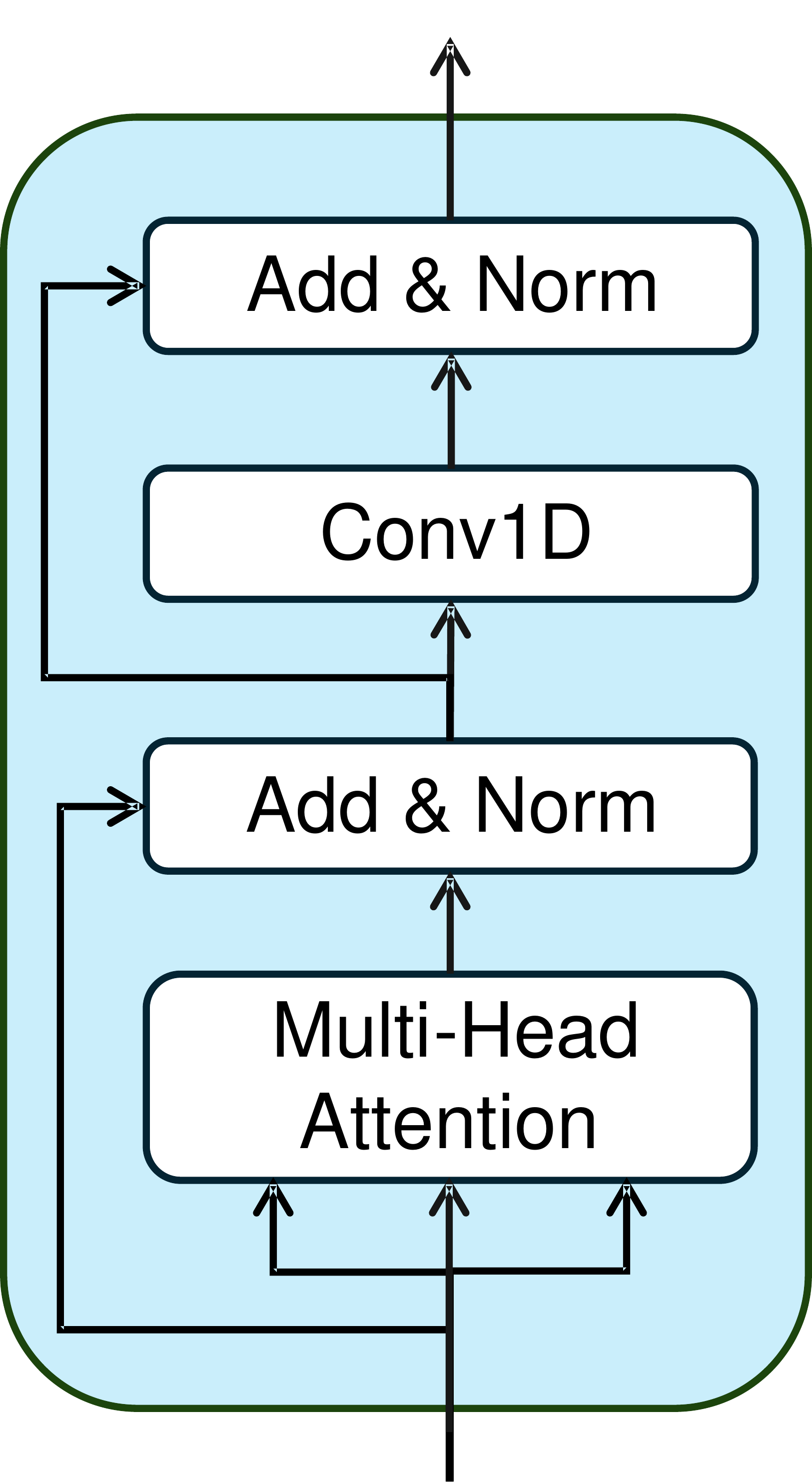}
                \caption{Feed Forward Transformer}
                \label{fig:fft}
            \end{subfigure}
        \end{subfigure}
     
    \end{subfigure}

    \caption{\textbf{StyleSpeech Architecture Diagram}: Figure~\ref{fig:style_speech} presents StyleSpeech Overview. Parameters within the layers marked by blue lines are frozen during LoRA training, while parameters along the green path remain trainable, marking the area for style adaptation. "Stage N fusion" denotes the location where style and phoneme fusion occurs. 
    Figure~\ref{fig:fusion} shows the Style Fusion Process.  Figure~\ref{fig:ape}: shows Structures of Acoustic Feature Encoder.  Figure~\ref{fig:fft}: shows the structure of FFT Blocks.}
\end{figure*}

\section{StyleSpeech}\label{sec:method}
In this section, we introduce the architecture of the StyleSpeech framework. Diverging from the conventional feed-forward structure, we introduce a novel Style Decorator structure designed specifically to adapt style features while preserving distinct phoneme features. We provide an overview of the entire system in Figure~\ref{fig:style_speech}, followed by detailed discussions of the individual components of StyleSpeech in the subsequent subsections.

\subsection{Acoustic Pattern Encoder}
Typical inputs for TTS task is a sentence, that consists of a sequence of words or characters, denoted as~\(X = (x_1, \ldots, x_n)\). In this work, we first transform the input sentence~\(X\) into sequences of phonemes~\(P\) and styles~\(S\), which we collectively refer to as "acoustic features", using Grapheme-To-Phoneme~(G2P) conversion.

The objective of the Acoustic Pattern Encoder~(APE) is to convert these acoustic features into sequences of acoustic embeddings. These are denoted as~\(H_P = (h_{P1}, \ldots, h_{Pn})\) for phonemes and \(H_S = (h_{S1}, \ldots, h_{Sn})\) for styles, making them more suitable for further processing in deep learning frameworks.

To achieve this transformation, we employ feed-forward Transformer~(FFT)~\cite{ren2019fastspeech} to convert input sequences to embedding. Specifically, each FFT block comprises a self-attention block paired with a 1D convolutional neural network~(1DCNN). The self-attention mechanism incorporates multi-head attention to capture positional relationships within the sequence. Additionally, the standard two-layer dense network in the Transformer~\cite{vaswani2017attention} is replaced with two 1DCNNs followed by ReLU activation. This modification enhances the model's ability to capture the close dependencies between adjacent hidden states, crucial for accurately representing the sequences of characters or phonemes and their corresponding mel-spectrograms in speech synthesis tasks~\cite{ren2019fastspeech}.

\subsection{Phoneme Duration Adaptor}
The duration of each phoneme in human speech varies from sentence to sentence. The \textit{Duration Adaptor} is employed to address the issue of length mismatch between the phoneme and spectrogram by adapting the length of each phoneme~\(n\) embedding~\(H_P\) or style embedding~\(H_S\) to match the length~\(m\) of the Mel-Spectrogram~\(Y\). 

Phoneme Duration Adaptor consists of two main components: the \textit{duration predictor} and the \textit{length regulator}. The duration predictor estimates the duration of each acoustic feature~\(L = \{l_1, \ldots, l_n\}, m = \sum_{i=0}^{N}{li}\). These predicted durations are then used to adjust the length of each acoustic embedding to adaptive embedding~\(H_L = {h_{l1},\ldots,h_{lm}}\). We adopt a similar setting to \cite{ren2020fastspeech} for training and deployment of the duration adaptor.

\subsection{Style Decorator}
Style integration is a process to incorporate style features~\(S = \{s_1, \ldots, s_n\}\) with the phoneme features~\(P = \{p_1, \ldots, p_n\}\) to create fused embeddings~\(H\). Similar to the Decorator Design pattern in object-oriented programming~(OOP), a structure that allows additional functionality to be dynamically added to an individual object without affecting its core behavior. Style Decorator is a structure designed to incorporate style features~\(S\) while preserving the distinct characteristics of exsiting acoustic features. The objectives of the Style Decorator are twofold: first, the adaptation process should not alter existing acoustic features such as phonemes; second, the style encoder model should be easily added or removed to incorporate or exclude style features from the TTS system.
Traditionally, integrating~\(S\) within a feed-forward TTS system, such as a variance adaptor~\cite{ren2020fastspeech},  requiring updating the model parameters~\(W_1 = W_0 + \Delta W\) by employing gradient descent across all layers. This method is often rigid and costly, as accommodating new style features typically requires extensive model re-training.

Drawing inspiration from the Lower-Rank Adaptation~\cite{hu2021lora} used in LLM and the style transfer method proposed in AutoVC~\cite{qian2019autovc}, we can we can approach the pre-trained TTS system as concrete LLM and each style feature as as a decorator for downstream task. Here, the modification~\(\Delta W\) can be approximated with trainable parameters~\(U\), which, in this study, are the trainable parameters within the FFT block that convert style acoustic patterns~\(S\) into style embedding~\(H_S\). The fusion of phoneme and style embeddings in the forward pass can be represented as:
\[H = WH_P+\Delta W(H_PH_S) = WH_P + UH_S\]

This approach offers two significant advantages. 
\textbf{Preservation of unique acoustic features}: the phoneme embeddings are not altered during this process, as the parameters~\(W\) remain fixed, and only~\(U\) is trained to adapt to new style features.
\textbf{Efficient Adaptation}:  Introducing a new style feature~\(S_1\) into a system already containing~\(S_0\) only requires updating the parameters~\(U_1\) specific to~\(S_1\). Unlike traditional feed-forward structures that requires comprehensive updates across all layers, including both existing and new parameters, the Style Decorator structure simplifies this to~\(\Delta W = U\), which is less resource-intensive and allows for quicker adaptation.

The fused embeddings~\(H\) is then processed through Mel-Spectrogram encoder that consisted of several layers of FFT to generate the Mel-Spectral embedding~\(H_M\). This is followed by a linear layer to map~\(H_M\) to the dimensions of a Mel-Spectrogram to produce the output, denoted by~\(Y'\). The model's training objective is to minimize the Mean Square Error~(MSE) Loss, ensuring that \(Y'\) closely aligns with the target output~\(Y\), which represents the actual speech in the frequency domain as a Mel-Spectrogram.

\subsection{Vocoder}
In this study, we use the Griffin-Lim algorithm-based vocoder~\cite{griffin1984signal} to transform the Mel-Spectrogram~\(Y\), back to its speech audio~\(A\). 
Specifically, the Griffin-Lim algorithm focuses on reconstructing the phase estimation~\(P(Y)\),  which indicates the position of each sinusoidal waveform within its cycle at each time frame represented in the Mel-Spectrogram~\(Y\). This phase estimation is iteratively updated based on the phase and complex-valued spectrogram~\(Y \cdot e^{iP(Y)}\) from the previous step, using the formula~\[P(Y^{(t+1)}) = P(Y^{(t)})Y^{(t)} \cdot e^{iP(Y^{(t)})}\]
Once convergence is achieved, the final complex-valued spectrogram~\(A'\) is computed as~\(A'=Y\cdot e^{iP(Y)}\), and the speech is reconstructed using the inverse Short Term Fourier Transform, \(A = ISFT(A')\). 
\begin{algorithm}[t]
\caption{StyleSpeech Pipeline}
\label{alg:pseduocode}
\SetAlgoLined
\SetAlgoNlRelativeSize{-1}
\DontPrintSemicolon
\LinesNumbered
\KwIn{$X$: Input Sentence}
\KwOut{A: Synthesised speech audio}
\SetKwProg{Procedure}{Procedure}{}{\KwRet{Speech Audio}}
\Procedure{StyleSpeech}{
    $(P, S) \gets \text{G2P}(X)$ \;
    $(H_P, H_S) \gets \text{APE}(P, S)$  \;
    $H_L \gets \text{DurationAdaptor}(H_P, H_S)$ \;
    $H \gets \text{StyleDecorator}(H_L, H_S)$ \;
    $Y \gets \text{MelSpectrogramEncoder}(H)$ \;
    $\text{A} \gets \text{Vocoder}(Y)$ \;
}
\end{algorithm}

\section{Experiments}
\subsection{Dataset}
In this study, we chose Chinese as our evaluation language due to its unique linguistic characteristics. Unlike English, Chinese uses characters to represent meanings rather than speech, resulting in the absence of a phonetic or syllabic writing system. To learn the pronunciation of Chinese characters, we must transcribe them into the Pinyin system, which contains phonemes and tones to describe the sound of each character. This poses a challenge in Chinese TTS systems because most applications treat Chinese characters or their Pinyin phonemes combined with tone symbols as a single acoustic pattern input, requiring a large number of embeddings for model training. 
However, our method encodes phonemes and tones as separate embeddings and then combines them using our style decorator structure. This strategy not only simplifies the model's architecture by reducing the number of embeddings required but also decreases the learning complexity.

We selected public Baker dataset~\cite{BakerDataset2020} to evaluate our method. The Baker dataset contains 10,000 high-quality voice recordings, all in 16-bit WAV format with a sampling frequency of 48kHz. These recordings are the work of a professional chinese voice actress aged between 20 and 30, with an elegant and optimistic vocal tone.
Pinyin phonemes serve as the phonemic input \(X\), while tone serves as the style input 
\(S\). We convert speech files into Mel-Spectrograms~\(Y\), with a frame size of 1024 and a hop length of 512. The dataset is split, allocating 4,000 sentences for training and 1,000 sentences for testing.

\subsection{Configuration}
StyleSpeech contains phoneme, style, and Mel-Spectrogram encoders, each with four FFT blocks. To ensure compatibility with the dimensional requirements of actual Mel-Spectrograms, the output of the Mel-Spectrogram embedding is transformed to 80-dimensional Mel-Spectrogram using through an 80-dimensional linear layer.

For optimization, we adjust the learning rate with a warm-up strategy from the Transformer model~\cite{vaswani2017attention}, setting dropout rates at 0.5 for FFT blocks and 0.1 for the length adaptor to prevent overfitting.

Furthermore, we conduct an ablation study to analyze the impact of integrating style embedding at different stages: 1) fusion before the length adaptor, 2) fusion after the length adaptor, and 3) fusion immediately before the linear layer. Please refer to Figure~\ref{fig:style_speech} for more details. 

We also explore how training methods impact TTS performance, comparing joint training of phonemic and style encoders with LoRA training, where the phonemic encoder is fixed and the style encoder and Mel-Spectrogram encoder are fine-tuned for new style features.

\subsection{Evaluation Metrics}
\begin{table*}[!t]
\centering

\begin{tabular}{lccccc}
\hline
\textbf{Model} & \textbf{WER~(\(\downarrow\))} & \textbf{WER-P~(\(\downarrow\))} & \textbf{WER-T~(\(\downarrow\))} & \textbf{MCD~(\(\downarrow\))} & \textbf{PESQ~(\(\uparrow\))} \\
\hline
FastSpeech & 0.419 $\pm$ 0.184 & 0.211 $\pm$ 0.128 & 0.342 $\pm$ 0.153 & 13.003 $\pm$ 4.081 & 1.054 $\pm$ 0.055 \\
\hline
\textbf{Joint Training} \\
\hline
StyleSpeech\_0 &\textbf{ 0.409 $\pm$ 0.179} & \textbf{0.195 $\pm$ 0.127} & \textbf{0.337 $\pm$ 0.149} & 12.826 $\pm$ 4.040 & 1.055 $\pm$ 0.056 \\
StyleSpeech\_1 & 0.972 $\pm$ 0.174 & 0.958 $\pm$ 0.175 & 0.439 $\pm$ 0.209 & \textbf{12.568 $\pm$ 3.865} & \textbf{1.333 $\pm$ 0.338} \\
StyleSpeech\_2 & 0.451 $\pm$ 0.180 & 0.265 $\pm$ 0.136 & 0.354 $\pm$ 0.147 & 15.091 $\pm$ 4.732 & 1.051 $\pm$ 0.053 \\
\hline
\textbf{LoRA Training} \\
\hline
StyleSpeech\_0 & \textbf{0.312 $\pm$ 0.156} & \textbf{0.220 $\pm$ 0.140} & \textbf{0.171 $\pm$ 0.112} & 12.843 $\pm$ 4.009 & 1.058 $\pm$ 0.059 \\
StyleSpeech\_1 & 0.388 $\pm$ 0.184 & 0.296 $\pm$ 0.173 & 0.223 $\pm$ 0.143 & \textbf{12.170 $\pm$ 3.958} & \textbf{1.087 $\pm$ 0.08}1 \\
StyleSpeech\_2 & 0.456 $\pm$ 0.189 & 0.263 $\pm$ 0.143 & 0.361 $\pm$ 0.154 & 14.977 $\pm$ 4.698 & 1.052 $\pm$ 0.053 \\

\hline
\end{tabular}%

\caption{Evaluation Results of TTS systems. WER-P and WER-T refer to Word Error Rate for phoneme and tone symbols, respectively. \textbf{(\(\downarrow\))} indicates that lower values are better, and \textbf{(\(\uparrow\))} indicates that higher values are better. The best-performing method for each metric within each training strategy is highlighted in \textbf{bold}.}

\label{tab:overall_results}
\end{table*}

\subsubsection{Quantitative Metric}\label{sec:quant_metric}
Initially, we use the TTS system to generate speech outputs. We employ Word Error Rate~(WER), Mel Cepstral Distortion~(MCD)~\cite{kubichek1993mel}, and Perceptual Evaluation of Speech Quality~(PESQ)~\cite{rix2001perceptual}, to quantitatively assess model's performance.

We assess the accuracy of synthesized speech using WER by first generating speech with a TTS system and then transcribing it through OpenAI's Whisper API~\cite{radford2023robust}. We compare these transcriptions to the original text, with a lower WER indicating better synthesis accuracy.  
In speech synthesis, we prioritize speech accuracy over written characters. Thus, we convert the transcript to Pinyin phonemes, such as "\{ni3 hao3\}," and separate these into phonemes "\{n, i, h, ao\}" and tones "\{0, 3, 0, 3\}" to calculate WER at the phoneme and style levels.


\begin{figure*}[ht]
    \centering
    \begin{subfigure}[t]{0.2\textwidth}
        \includegraphics[width=\textwidth]{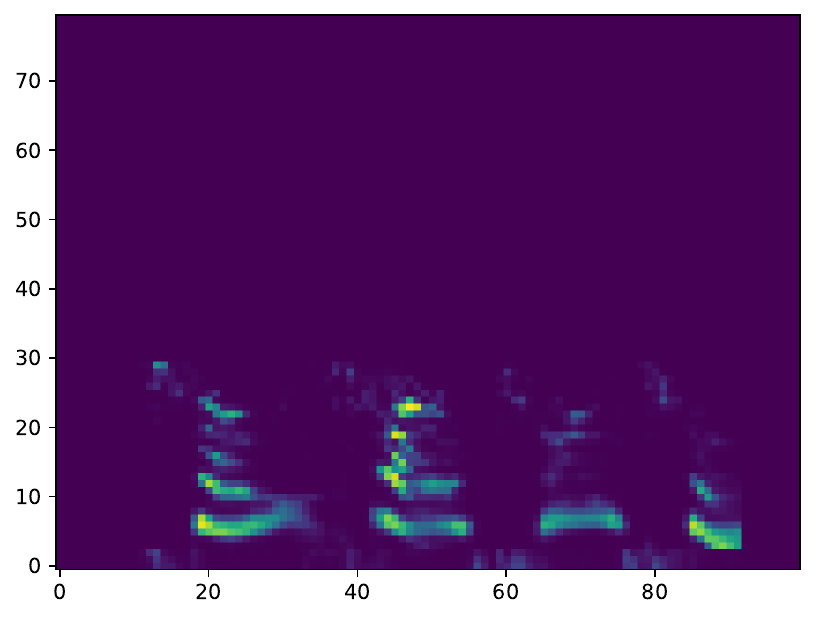}
        \caption{FastSpeech}
        \label{fig:fastspeech}
    \end{subfigure}
    \begin{subfigure}[t]{0.2\textwidth}
        \includegraphics[width=\textwidth]{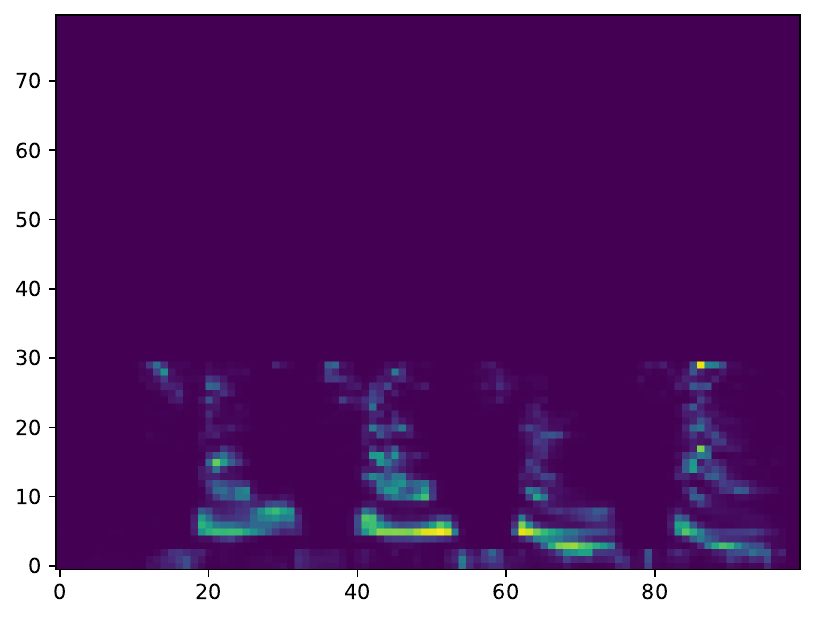}
        \caption{StyleSpeech 0 joint}
        \label{fig:s0}
    \end{subfigure}
    \begin{subfigure}[t]{0.2\textwidth}
        \includegraphics[width=\textwidth]{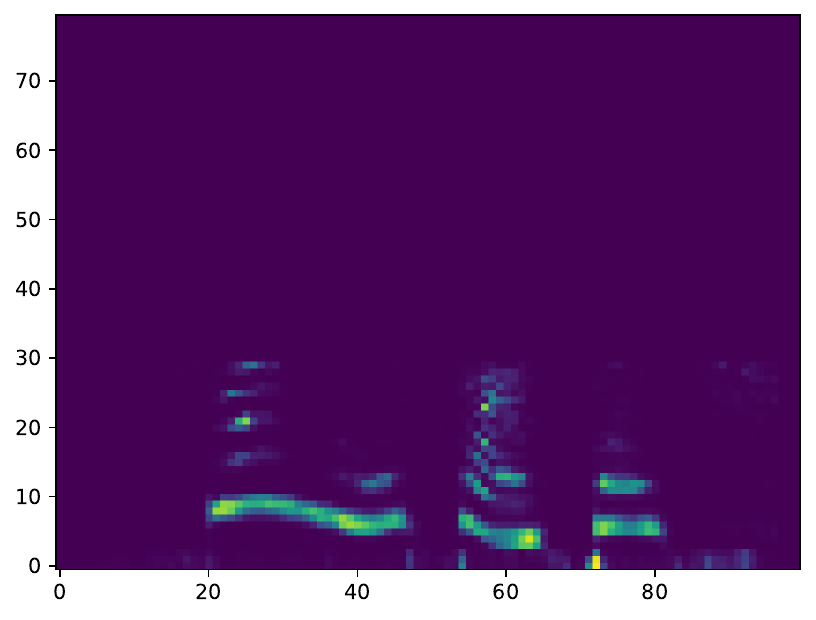}
        \caption{StyleSpeech 1 joint}
        \label{fig:s1}
    \end{subfigure}
    \begin{subfigure}[t]{0.2\textwidth}
        \includegraphics[width=\textwidth]{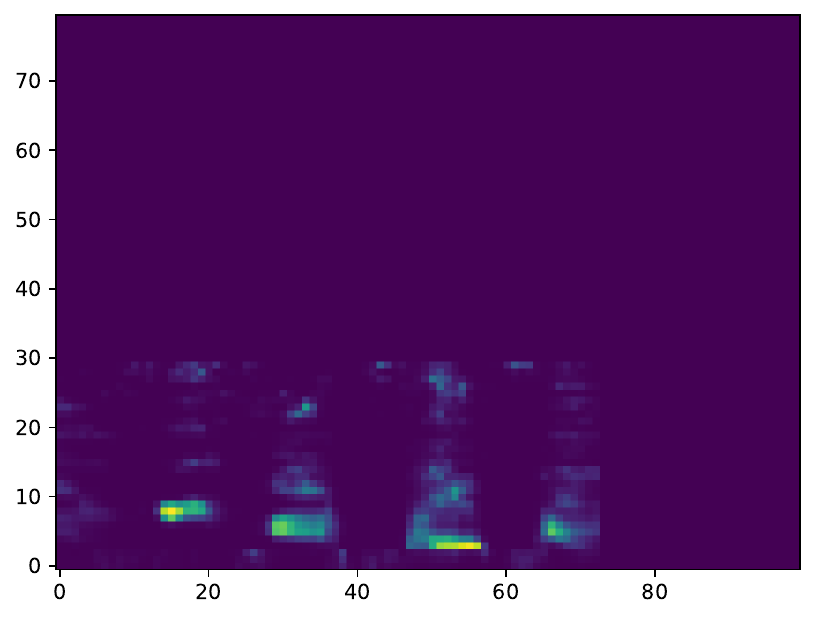}
        \caption{StyleSpeech 2 joint}
        \label{fig:s2}
    \end{subfigure}
    \begin{subfigure}[t]{0.2\textwidth}
        \includegraphics[width=\textwidth]{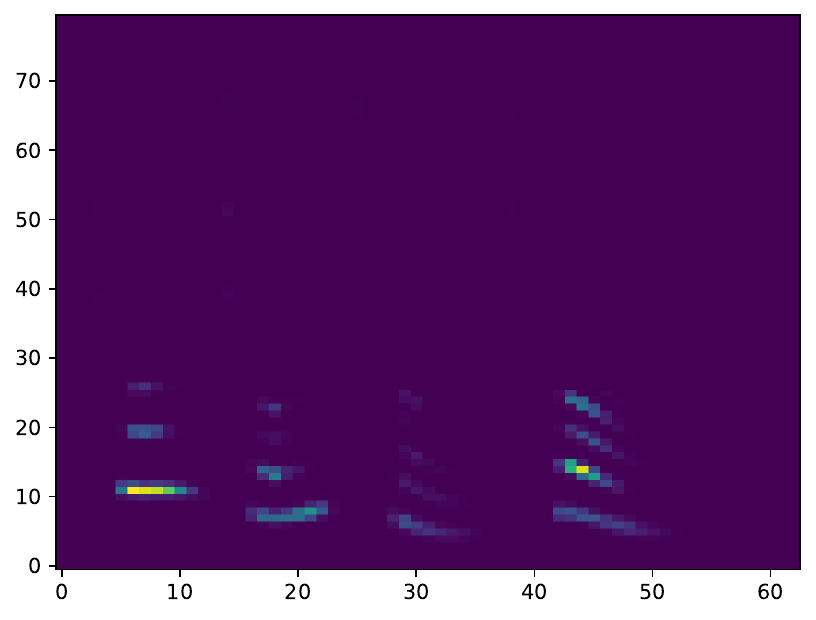}
        \caption{Ground Truth}
        \label{fig:real}
    \end{subfigure}
    \begin{subfigure}[t]{0.2\textwidth}
        \includegraphics[width=\textwidth]{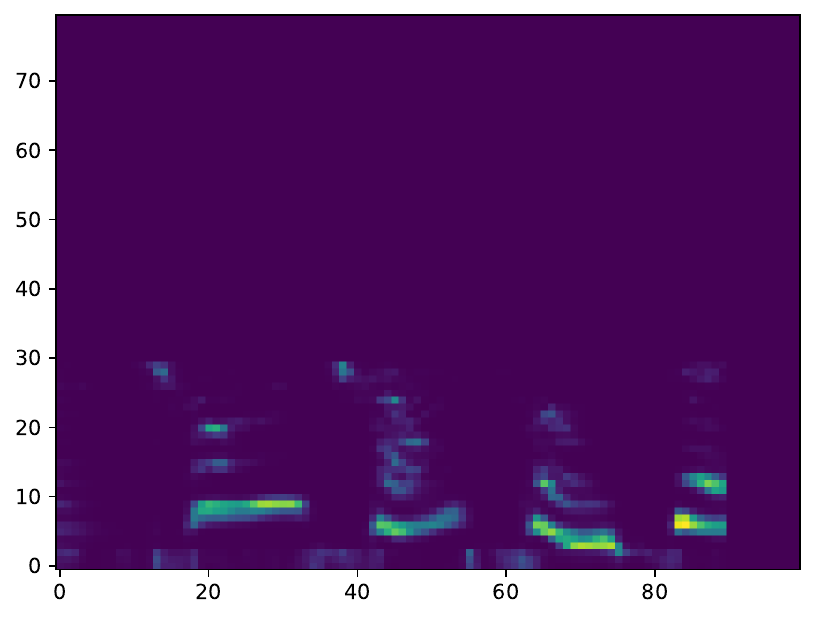}
        \caption{StyleSpeech 0 LoRA}
        \label{fig:s0f}
    \end{subfigure}
    \begin{subfigure}[t]{0.2\textwidth}
        \includegraphics[width=\textwidth]{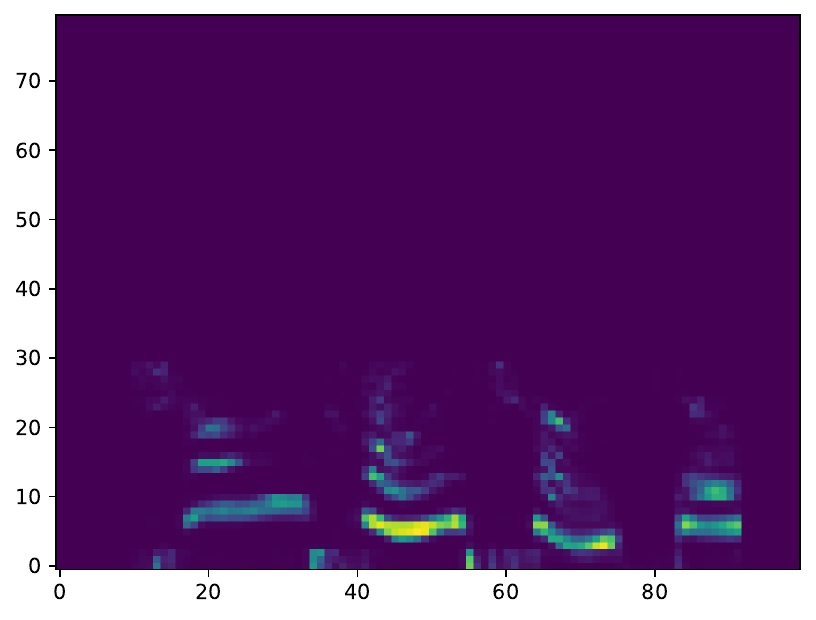}
        \caption{StyleSpeech 1 LoRA}
        \label{fig:s1f}
    \end{subfigure}
    \begin{subfigure}[t]{0.2\textwidth}
        \includegraphics[width=\textwidth]{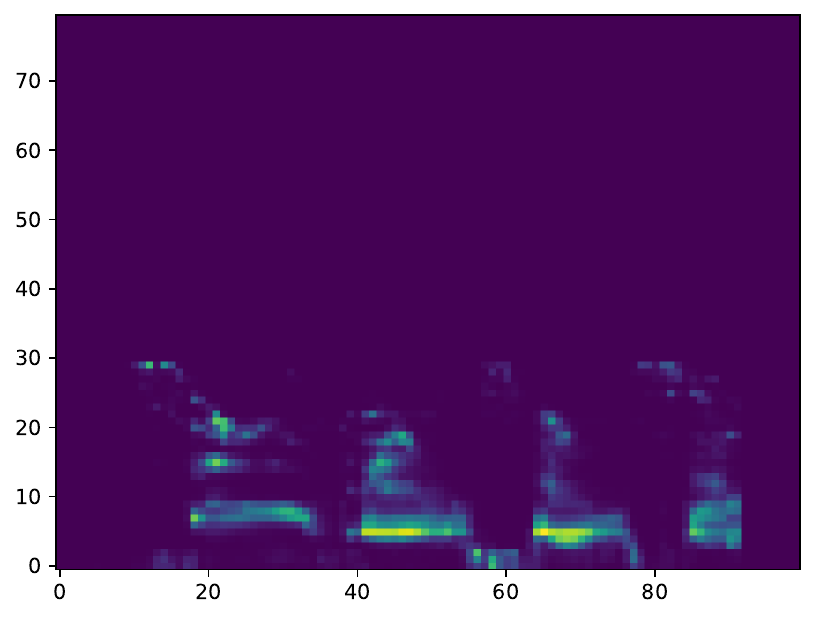}
        \caption{StyleSpeech 2 LoRA}
        \label{fig:s2f}
    \end{subfigure}
    \caption{Mel-Spectrogram diagram produced by various TTS systems when synthesizing the Pinyin phoneme~\textit{Chong} with different tones, \textit{Chong1}, \textit{Chong2}, \textit{Chong3}, and \textit{Chong4}. Since the values for the ground truth diagram above 30 frequency are inactive, we have cropped all values above 30 in the synthesized speech for easier comparison.}
    \label{fig:melspectro}
\end{figure*}

\subsubsection{LLM-Guided MOS}
We introduce LLM-MOS to qualitatively rate synthesized speech quality on a scale of 1 to 5, combining the precision of quantitative metrics with the nuanced insight of an LLM to reduce subjectivity found in traditional MOS evaluations. This involves synthesizing speech from different TTS systems, calculating metric thresholds based on percentile rankings, and assigning a 1-5 rating according to these thresholds. Each speech's overall quality rating is then derived by averaging its WER, MCD, and PESQ scores, providing a comprehensive and straightforward evaluation method for TTS performance.

\section{Results and Discussion}

\subsection{Imapact of Style Decorator}
In Table~\ref{tab:overall_results}, we present the impact of the Style Decorator on various quantitative metrics, as detailed in Section~\ref{sec:quant_metric}. The best-performing method for each metric within each training strategy is highlighted in \textbf{bold}. Table~\ref{tab:statistically_significant_test} shows the statistically significant tests compared with the baseline model. WER-P stands for \textbf{P}noneme-level WER and WER-T stands for \textbf{T}one-level WER.

The data in Table~\ref{tab:overall_results} vividly shows that StyleSpeech outperforms the baseline model across all metrics, particularly with significant improvements in both WER and MCD metrics. Notably, StyleSpeech achieves over a 10\% enhancement in overall WER and a 15\% improvement in tone-level WER. Since the phoneme-level improvements are minimal, it suggests that the major gains in overall WER primarily stem from the system’s enhanced ability to accurately synthesize tone styles. This underscores the efficacy of the Style Decorator structure and its robust capacity to adapt style features.

Conversely, improvements in PESQ display a different trend, differing from those observed for WER and MCD, which generally relate to fusion at stage 0. Improvements in PESQ are more closely associated with stage 1 fusion, particularly after the length adaptor. This distinction indicates that the acoustic effects of different fusion stages are varied, influencing the synthesized output in distinct manners, which will be explored further in Section~\ref{sec:fuse_step}.

The LLM-MOS metric in Table~\ref{tab:llm-mos-ratings} aligns with these performance trends, demonstrating substantial enhancements: around 21\% in WER, 10\% in MCD, and an impressive 82\% in PESQ. This metric provides a clear and concise representation of performance comparisons.

Figure~\ref{fig:melspectro} showcases Mel-Spectrogram diagrams produced by different TTS systems for the same phoneme across varying tones. Figure~\ref{fig:real} illustrates the Mel-Spectrogram for the ground truth speech. When StyleSpeech is applied in conjunction with the LoRA training configuration, as depicted in Figures~\ref{fig:s0f} and \ref{fig:s1f}, it exhibits clearer boundary between frequency bands, closely mirroring the characteristics of the ground truth speech. In comparison, the diagrams generated by FastSpeech contain denser frequency layers, leading to extraneous noises in the final speech production and a less accurate synthesis of the input phoneme.


\begin{table}[t]
    \centering
    \resizebox{0.5\textwidth}{!}{%
    \begin{tabular}{c c c c c}
    \toprule
    \bf Strategy & \bf Metric & \multicolumn{3}{c}{\bf Method} \\
    \midrule
    \multirow{6}{*}{\bf Joint} & \multirow{2}{*}{\bf WER P-Value} &\textbf{ \textbf{StyleSpeech 0}} & \textbf{StyleSpeech 1} & \textbf{StyleSpeech 2} \\
                               && 0.23 & \textbf{0.00} & \textbf{0.00} \\
    \cmidrule[0.5pt](lr){2-5}
                               & \multirow{2}{*}{\bf MCD P-Value} & \textbf{StyleSpeech 0} & \textbf{StyleSpeech 1} & StyleSpeech 2 \\
                               && \textbf{0.00} & \textbf{0.00} & \textbf{0.00} \\
    \cmidrule[0.5pt](lr){2-5}
                               & \multirow{2}{*}{\bf PESQ P-Value} & StyleSpeech 0 & \textbf{StyleSpeech 1} & StyleSpeech 2 \\
                               && 0.69 & \textbf{0.00} & 0.27 \\
    \midrule
    \multirow{6}{*}{\bf LoRA}  & \multirow{2}{*}{\bf WER P-Value} & \textbf{StyleSpeech 0} & \textbf{StyleSpeech 1} & \textbf{StyleSpeech 2} \\
                               && \textbf{0.00} & \textbf{0.00} & \textbf{0.00} \\
    \cmidrule[0.5pt](lr){2-5}
                               & \multirow{2}{*}{\bf MCD P-Value} & \textbf{StyleSpeech 0} & \textbf{StyleSpeech 1} & StyleSpeech 2 \\
                               && \textbf{0.00} & \textbf{0.00} & 0.30 \\
    \cmidrule[0.5pt](lr){2-5}
                               & \multirow{2}{*}{\bf PESQ P-Value} & StyleSpeech 0 & \textbf{StyleSpeech 1} & StyleSpeech 2 \\
                               && 0.10 & \textbf{0.00} & 0.30 \\
    \bottomrule
    \end{tabular}
    }
    \caption{Statistically Significant Tests. We find that 12 out of 18 comparisons are significant~($p \leq 0.05$), with results shown in \textbf{bold}.}
    \label{tab:statistically_significant_test}
\end{table}

\subsection{Joint versus LoRA training}
In terms of how training methodologies impact model performance, LoRA training typically results in greater enhancements compared to joint training. This is evident in stage 1 fusion results, where LoRA training maintains lower Word Error Rates~(WER) at 0.296, significantly outperforming joint training, which sees WER soar to 0.958. The superior performance of LoRA training is due to its ability to freeze phonemic parameters during the training process, which helps preserve the unique characteristic inside phonetic features.

In contrast, joint training tends to merge style and phoneme embeddings, making it challenging for the model to accurately interpret and align them with the corresponding Mel-Spectrogram. This blending can lead to misrepresentations in the output and a loss of distinguishable details between different phonemes. As demonstrated in Figures~\ref{fig:s0} and ~\ref{fig:s1}, speech generated under joint training exhibit blurrier frequency boundaries compared to those from LoRA training.
By treating style feature as an additive layer or 'decorator' of phoneme feature, LoRA training enables the seamless integration of new style variations without disrupting the phonemic structure, ensuring that each phoneme retains its integrity and is distinctly recognizable. Conversely, the increased complexity of the learning task in joint training can confuse the model and make parameter tuning more challenging.

\subsection{Impact of Fusion Stage}\label{sec:fuse_step}
The impact of the fusion stage between phoneme and style embeddings on a method's performance is significant. The fusion stage occurring at lower layers tends to influence the output more significantly, whereas those closer to higher layers have a lesser impacts.

StyleSpeech produces the most accurate speech when fusion occurs at an early stage, before the length adaptor. This is demonstrated in both Table~\ref{tab:overall_results} and Table~\ref{tab:llm-mos-ratings}, where StyleSpeech 0 has the lowest WER score under both joint and LoRA training cases. This early-stage fusion combines phoneme and style embeddings, allowing the length adaptor to adjust the duration of each phoneme based on these hybrid embeddings. Such synchronization enhances control over the synthesized speech, thereby improving its accuracy.

Conversely, fusion at stage 1, presents a paradox. 
Although the perceptual quality of the speech is improved, the accuracy diminishes. This decrease in accuracy occurs because the style embeddings do not influence the phoneme duration predictions, leaving the durations solely determined by phoneme embeddings. This approach fails to capture the correct timing of speech, often resulting in outputs that are unrealistic and indistinguishable. For example, in Figure~\ref{fig:s1}, the first and second phonemes merge, creating a single phoneme. However, fusing post-length adaptor does enhance the distinction between phoneme and style features, leading to higher perceptual quality, as shown in Figure~\ref{fig:s1} where frequency boundaries are more defined compared to the early-stage fusion version in Figure~\ref{fig:s0}.

Fusion at the final stage appears to pose minimal influence on performance.  By this stage, the hidden embeddings are already saturated with strong acoustic and tonal characteristics. Thus, additional fusion at this point does not substantially modify speech, leading to negligible effects on the output.
\begin{table}[t]
\centering
\resizebox{0.5\textwidth}{!}{
\begin{tabular}{lcccc}
\hline
\textbf{Model} & \textbf{WER} & \textbf{MCD} & \textbf{PESQ} & \textbf{Overall} \\
\hline
FastSpeech & 3.27 $\pm$ 1.24 & 3.08 $\pm$ 1.39 & 2.55 $\pm$ 1.31 & 2.99 $\pm$ 1.01 \\
\hline
\textbf{Joint Training} \\
\hline
StyleSpeech\ 0 & \textbf{3.34 $\pm$ 1.22} & 3.13 $\pm$ 1.37 & 2.62 $\pm$ 1.29 & \textbf{3.04 $\pm$ 0.98} \\
StyleSpeech\ 1 & 1.00 $\pm$ 0.05 & \textbf{3.22 $\pm$ 1.34} & \textbf{4.64 $\pm$ 0.72} & 3.00 $\pm$ 0.65 \\
StyleSpeech\ 2 & 3.06 $\pm$ 1.22 & 2.54 $\pm$ 1.44 & 2.45 $\pm$ 1.35 & 2.68 $\pm$ 1.06 \\
\hline
\textbf{LoRA Training} \\
\hline
StyleSpeech\ 0 & \textbf{4.03 $\pm$ 1.02} & 3.13 $\pm$ 1.37 & 2.80 $\pm$ 1.19 & 3.31 $\pm$ 0.91 \\
StyleSpeech\ 1 & 3.51 $\pm$ 1.17 & \textbf{3.36 $\pm$ 1.35} & \textbf{3.52 $\pm$ 1.06} & \textbf{3.48 $\pm$ 0.88} \\
StyleSpeech\ 2 & 3.02 $\pm$ 1.26 & 2.55 $\pm$ 1.43 & 2.44 $\pm$ 1.33 & 2.66 $\pm$ 1.08 \\
\hline
\end{tabular}
}
\caption{LLM-Guided MOS Ratings. A higher rating indicates better speech quality. The best-performing method for each metric within each training strategy is highlighted in \textbf{bold}.}
\label{tab:llm-mos-ratings}
\end{table}

\section{Conclusion}
In this study, we present StyleSpeech, a novel TTS system that can accurately synthesize human speech and efficiently adapt to style features while preserving phoneme features. Furthermore, we developed an automated qualitative metric, LLM-MOS, designed to provide an objective evaluation of TTS systems relative to others, ensuring a more equitable assessment. 

The limitations of this study include a focus on a single language, which may limit the generalizability of our findings. Expanding to multiple languages can help assess the transferability of our methods. Additionally, our use of simple additive fusion techniques may restrict the TTS system's performance. Therefore, further research could explore advanced methods, such as Mixture of Experts (MOE), and refine LLM evaluation prompts to enhance both the system's performance and the precision of our evaluations. Our future research will primarily focus on incorporating a broader range of style features and enhancing fusion strategies to further advance the capabilities of the TTS system.
\bibliographystyle{ACM-Reference-Format}
\bibliography{reference}

\end{document}